# Gamma irradiated nanostructured NiFe$_2$O$_4$: Effect of γ-photon on morphological, structural, optical and magnetic properties

Sapan Kumar Sen[a,*], Majibul Haque Babu[b], Tapash Chandra Paul[c], Md. Sazzad Hossain[d], Mongur Hossain[e], Supria Dutta[f], M. R. Hasan[g], M. N. Hossain[h], M. A. Matin[h], M. A. Hakim[h], Parimal Bala[c]

[a]Institute of Electronics, Atomic Energy Research Establishment, Bangladesh Atomic Energy Commission, Dhaka-1349, Bangladesh
[b]Basic Science Division, World University of Bangladesh, Dhaka-1205, Bangladesh.
[c]Department of Physics, Jagannath University, Dhaka-1100, Bangladesh.
[d]Department of Physics, University of Dhaka, Dhaka-1000, Bangladesh.
[e]Hunan Key Laboratory of Two-Dimensional Materials, State Key Laboratory for Chemo/Biosensing and Chemometrics, College of Chemistry and Chemical Engineering, Hunan University, Changsha, 410082 China.
[f]Ministry of Education, Government of the People's Republic of Bangladesh, Dhaka, Bangladesh.
[g]Materials Science Division, Atomic Energy Center, Dhaka, Bangladesh Atomic Energy Commission, Dhaka-1000, Bangladesh.
[h]Department of Glass & Ceramic Engineering, Bangladesh University of Engineering & Technology, Dhaka-1000, Bangladesh.

## Abstract

The current manuscript highlights the preparation of NiFe$_2$O$_4$ nanoparticles by adopting sol-gel auto combustion route. The prime focus of this study is to investigate the impact of γ-irradiation on the microstructural, morphological, functional, optical and magnetic characteristics. The resulted NiFe$_2$O$_4$ products have been characterized employing numerous instrumental equipments such as FESEM, XRD, UV–visible spectroscopy, FTIR and PPMS for a variety of γ-ray doses (0 kGy, 25 kGy and 100 kGy). FESEM micrographs illustrate the aggregation of ferrite nanoparticles in pristine NiFe$_2$O$_4$ product having an average particle size of 168 nm and the surface morphology is altered after exposure to γ-irradiation. XRD spectra have been analyzed employing Rietveld method and the results of the XRD investigation reveal the desired phases (cubic spinel phases) of NiFe$_2$O$_4$ with observing other transitional phases. Several microstructural parameters such as bond length, bond angle, hopping length etc. have been determined from the analysis of Rietveld method. This study reports that the γ-irradiations demonstrate a great influence on optical bandgap energy and it varies from 1.80 and 1.89 eV evaluated via K-M function. FTIR measurement depicts a proof for the persistence of Ni-O and Fe-O stretching vibrations within the respective products and thus indicating the successful development of NiFe$_2$O$_4$. The saturation magnetization (M$_S$) of pristine Ni ferrite product is noticed to be 28.08 emu/g. A considerable increase in M$_S$ is observed in case of low γ-dose (25 kGy) and a decrement nature is disclosed after the result of high dose of γ-irradiation (100kGy).

**\*Corresponding author:** sapansenphy181@gmail.com; ORCID ID: 0000-0001-5086-2758

# 1 Introduction

In recent years, nanocrystalline spinel ferrites have been investigated immensely owing to potential applications in chemical sensors, microwave absorbers, permanent magnets, high density recording systems, ferrofluid technology, biomedical, imaging and high-frequency device applications [1][2][3]. Importantly, they possess excellent both electrical and magnetic properties, which are extremely sensitive to the many factors such as chemical composition, synthesis method, annealing temperature or temperature treatment, and cation distribution at tetrahedral (A) and octahedral [B] sites [3]. Among the various spinel ferrites, however, $NiFe_2O_4$ (nickel ferrite) has drawn an optimum attention recently due to various applications in gas sensors [4], spintronics [5], microwave absorption [6], catalyst [7], lithium-ion batteries [8], hydrogen production [9], even in biomedicine [10] etc.. As more and more considerations have been devoted keenly to the nano-sized magnetic materials for inherent their unique properties compared to their bulk counterparts, so the scientific engrossment on nano-sized $NiFe_2O_4$ is on the expanding in the research community. In this direction, the magnetism of $NiFe_2O_4$ is predominantly intriguing due to its substantial saturation magnetization and unique magnetic structures. In general, $NiFe_2O_4$ belongs to an inverse spinel structure with $Ni^{2+}$ ions on octahedral B sites (denoted as $O_h$-sites) and $Fe^{3+}$ ions on both of the tetrahedral A (denoted as $T_d$-site and $O_h$-sites) sites equally [11]. This is typically maintained by the formation energy in favor of the reverse spinel rather than spinel structure [12]. It is also found to have the mixed spinel structure with the inverse one, i.e., some $Ni^{2+}$ ions may occupy the $T_d$- site [11][13][14].

A general formula for a nickel ferrite structure is $(Ni_{1-x}Fe_x) [Ni_xFe_{2-x}]O_4$, where x is the degree of inversion. According to the crystal field theory (CFT), magnetic moments are rising from the local moments of the $Ni^{2+}$ with $3d^8$ as well as $Fe^{3+}$ with $3d^5$ electrons. Significantly, the net magnetization comes from the $Ni^{2+}$ ($O_h$-sites) cations alone ($\sim 2\mu B$), while $Fe^{3+}$ moments ($\sim 5\mu B$) in a high spin state for both $O_h$- and $T_d$-sites are antiparallel and abandon with each other [11]. However, the modification in physical properties (i.e structural and magnetic properties) of nanoferrites can be justified by instigating radiation damage by means of swift heavy ions[15], laser beam [16], proton [17] and gamma radiation [18]. Lately, several irradiation systems are implemented in the cutting-edge world because it is a striking tool for modifying the physical properties of nanoparticles in the research and development in commercial applications and industrial technologies like aeronautical and satellite communication, pollution control, material development, security systems and so on [15][19][20]. Upon electromagnetic nature, penetrating power, very short wavelength and all medium propagation property, the gamma ray ($\gamma$) is considered as ionizing radiation [21]. So, the present inquiry deals with an interaction of $\gamma$ with $NiFe_2O_4$ which can tune its physical properties since it is known to generate controlled defects of various types such as point, cluster, and columnar defects in the materials [22]. Besides, irradiation with γ rays contain the plausibility of dislocation of $Fe^{3+}$ ions from tetrahedral A sites to $Fe^{2+}$ ions at octahedral B sites [19].

Several research works have been focused on different preparation routes to synthesize $NiFe_2O_4$ nanoparticles, such as co-precipitation [23], sol-gel [24], spray pyrolysis [25], mechanical activation [26], hydrothermal method [27], high energy ball milling [28], etc..Concurrently, to improve the physical properties of nanoferrites using γ rays, various methods have been employed to synthesize ferrite nanoparticles. Recently, Raut et al. [19] synthesized $ZnFe_2O_4$ using sol–gel auto-combustion technique and reported that saturation magnetization and magneton number increased by γradiation dose of 50 and 100 kGy. Raut et al.[29] also prepared $CoFe_2O_4$ by sol-gel auto combustion with total radiation doses of 50 and 100 kGy and showed that lattice parameters decline with Coercivity ($H_c$), Remanence magnetization ($M_r$) and anisotropy field ($H_k$) shrinkages as a function of γ radiation doses. Based on the above survey, till now, there is no study on the impact of γ radiations on $NiFe_2O_4$ nanoparticles synthesized by sol-gel auto combustion method. Therefore, the aim of the research work is to synthesize and investigate some of the physical parameters of $NiFe_2O_4$ before and after γ-irradiation.

## 2 Experimental sections

### 2.1 Materials

In this experiment, Nickel Nitrate Hexahydrate ($Ni(NO_3)_2.6H_2O$), Ferric Nitrate Nanohydrate ($Fe(NO_3)_3.9H_2O$), Citric Acid ($C_6H_8O_7.H_2O$), Ammonium Hydroxide ($NH_4OH$) were used for the sample preparation. The samples were made by mixing the compositions ($Ni(NO_3)_2.6H_2O$, $Fe(NO_3)_3.9H_2O$, $C_6H_8O_7.H_2O$ and $NH_4OH$) together. The required materials were employed as received.

### 2.2 Synthesis route

Sol-Gel Auto-Combustion route has been employed to prepare the $NiFe_2O_4$ nanoparticles as a function of γ-irradiation for 0, 25 and 100 kGy doses involving the metal nitrates of the ingredient components as raw materials in citric acid matrix. The synthesis route has been explained as follows: the metal nitrates and citric acid were dissolved in an appropriate amount to keep the molar ratio of metal ions and used citric acid to1:1. A little quantity of ammonia was gradually dissolved into the starting solution to balance pH = 7 and stabilize the nitrate-citrate sol. The obtained precursor aqueous solution was stirred vigorously with a magnetic stirrer at 60 °C. After that the sol was placed into a tray and heated gradually to 120 °C to change into an extremely viscous brown gel. The gel was slowly heated to 250 °C in order to attain dried gel and after few minutes, the gel started to completely burnt to form a crisp powder. The probable chemical reaction during the synthesis of $NiFe_2O_4$ products is given as:

$$Ni(NO_3)_2 + 2Fe(NO_3)_3 + \frac{20}{9}C_6H_0O_7 = NiFe_2O_4 + \frac{40}{3}CO_2 + \frac{80}{9}H_2O + 4N_2$$

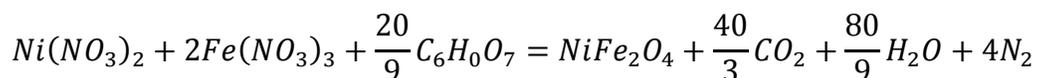

### 2.3 Co-60 Gamma irradiation

The synthesized NiFe$_2$O$_4$ products have been subjected to γ-irradiation originated from a 60Co source with various doses (25 kGy and 100 kGy) at Institute of Food and Radiation Biology (IFRB), Atomic Energy Research Establishment, Bangladesh Atomic Energy Commission, Dhaka, Bangladesh. The activity of 60Co source at the time of exposure is 12.17kGy/h and liquid phase dosimetry arrangement (Ericcerous) has been employed to enumerate the γ-dose rate.

### 2.4 Characterization techniques

The morphological investigation of the products has been acquired by Field Emission Scanning Electron Microscopy (FESEM) (JEOL JSM-7600F, USA). X-ray Diffraction (XRD) (model X′PertPRO XRD Philips PW3040, Netherlands) has been applied on the resulted NiFe$_2$O$_4$ and γ-irradiated NiFe$_2$O$_4$ to identify the phases and crystallinity. XRD structure is equipped with a Cu-Kα radiation (λ = 1.5404 Å) maintained in the range of 2θ from 15° to 90°. XRD spectra have been investigated through the Rietveld technique as implemented in Fullporf software. To examine the functional groups associated with ferrite samples, Fourier Transform Infrared Spectroscopy (FTIR) equipment has been employed during this experiment. Physical Properties Measurement System (PPMS), Quantum Design Dyna Cool at ambient conditions has been used to measure the magnetic hysteresis loop on the obtained products as well as different magnetic parameters such as saturation magnetization ($M_s$), remnant magnetization ($M_r$) and coercive field ($H_c$). DRS informations has been recorded by a UV–Vis device (Model: PerkinElmer UV–Vis–NIR Spectrometer Lambda 1050) at a wavelength of 200–800 nm.

### 3 Results and discussion

### 3.1 Surface morphology properties

Fig. 1 reveals the FESEM images of pristine and γ-irradiated NiFe$_2$O$_4$ with comparatively low (25kGy) and high (100 kGy) doses of γ radiations. It is clearly seen in the FESEM images that the surface morphology of NiFe$_2$O$_4$ is being altered with total dose of γ radiation. The same outcomes have been detected in the previous literature [30][31][32]. According to the FESEM images [Fig. 1(a,d)], the highest average particle size is pronounced 168 nm for pristine sample. It is seen that result of employing a low γ dose, the surface disrupts and making smaller particles due to the lattice vibration, atomic displacement and local heating effects [33]. The disintegration of particles can also be attributed to induced compressive stress which causes intrinsic defect recombination or reordering of initially disordered phase and aids the particles to regain their shapes [34][35]. In addition, at a low dose, γ photon has a higher possibility of interaction with the materials and then photoelectric absorption dominates, resulting in the decrement of average particle size of NiFe$_2$O$_4$ from 168 to 135 nm [36]. Conversely, in case of high dose (100 kGy), the average particle size is slightly increased from 135 to 140 nm as shown in Fig. 1(c,f). The swelling of average particle size may be occurred due to the decrease of photoelectric absorption [37]. It is well known that particles amalgamate at higher γ dose owing to no electron transfer, thereby, to make sure increment of average particle size [38]. A hypothetical growth mechanism of NiFe$_2$O$_4$ particles in pristine and $\gamma$-irradiated samples with different γ doses is depicted in Fig. 2.

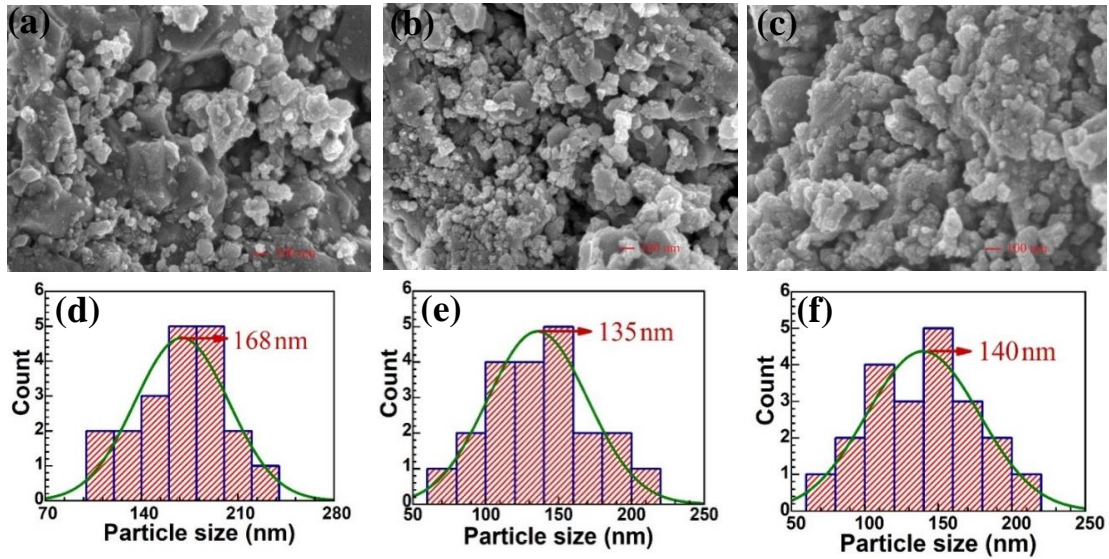

**Fig. 1** The FESEM images and particle size distribution of NiFe$_2$O$_4$ samples: (a, d) for 0 kGy, (b, e) for 25kGy, and (c, f) for 100 kGy doses.

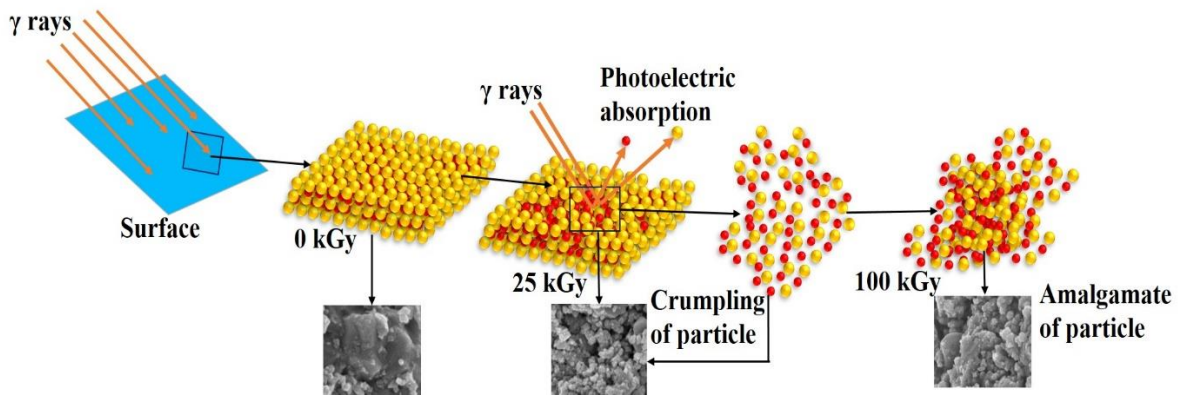

**Fig. 2** A plausible growth mechanism of pristine and $\gamma$-irradiated NiFe$_2$O$_4$ samples.

### 3.2 Structural properties

Fig. 3 depicts the Rietveld refinement (R.R) of the structure by XRD data is processed using Fullprof suite software. By a meticulous R.R analysis, a duel common inverse spinel cubic structure (Fd$\bar{3}$m) and hematite phase (R$\bar{3}$c) is firmly confirmed for the pristine and irradiated samples. It can be evidently indexed to the face centered cubic (fcc) structure of NiFe$_2$O$_4$ (JCPDS Card No.10-0325) [39] as shown in Fig. 3. On the other hand, a nickel phase (Fm$\bar{3}$m) has appeared for only the irradiated samples. From Fig. 3, it can be seen that the most prominent intense peak is manifested at $2\theta$ = 35.74° corresponding to the (311) plane of NiFe$_2$O$_4$ for pristine sample. After that, the place of (311) plane orientation has been changed from $2\theta = 35.75°$ to $35.79°$ by applying gamma radiation. This may be attributed to the formation of $Fe^{3+}$ ions in the place of $Fe^{2+}$ ions. It is also noticeable that the intensity of (3 1 1) peak increases while (400) peak intensity decline for applying 25 kGy gamma radiation. Conversely, owing to 100 kGy gamma radiation, the intensity of (400) peak rises sharply when (311) peak falls drastically. Such results, the shift of the peak intensity is in good agreement with that reported by several authors and they ascribed that to the lattice distortion arisen after irradiation [40][41].

To know the lattice parameters, we have done the R.R analysis using the FullProf software. The quality of the R.R is determined by means of a set of conventional statistical parameters. Typically, the quality of the R.R can be determined using several statistical parameters such as goodness of fit ($\chi^2$), weight Profile R-factor ($R_{wp}$) and expected R-Factor ($R_{exp}$), whereas $R_{wp}$ comparations the adjusted data with the experimental data, $R_{exp}$ estimates the quality of the experimental data and $\chi^2 = \left(R_{wp}/R_{exp}\right)^2$ [42]. During the R.R process, $\chi^2$ initiates with a maximum value when the goodness of fit model is poor and decreases as the fit data matches the experimental data.

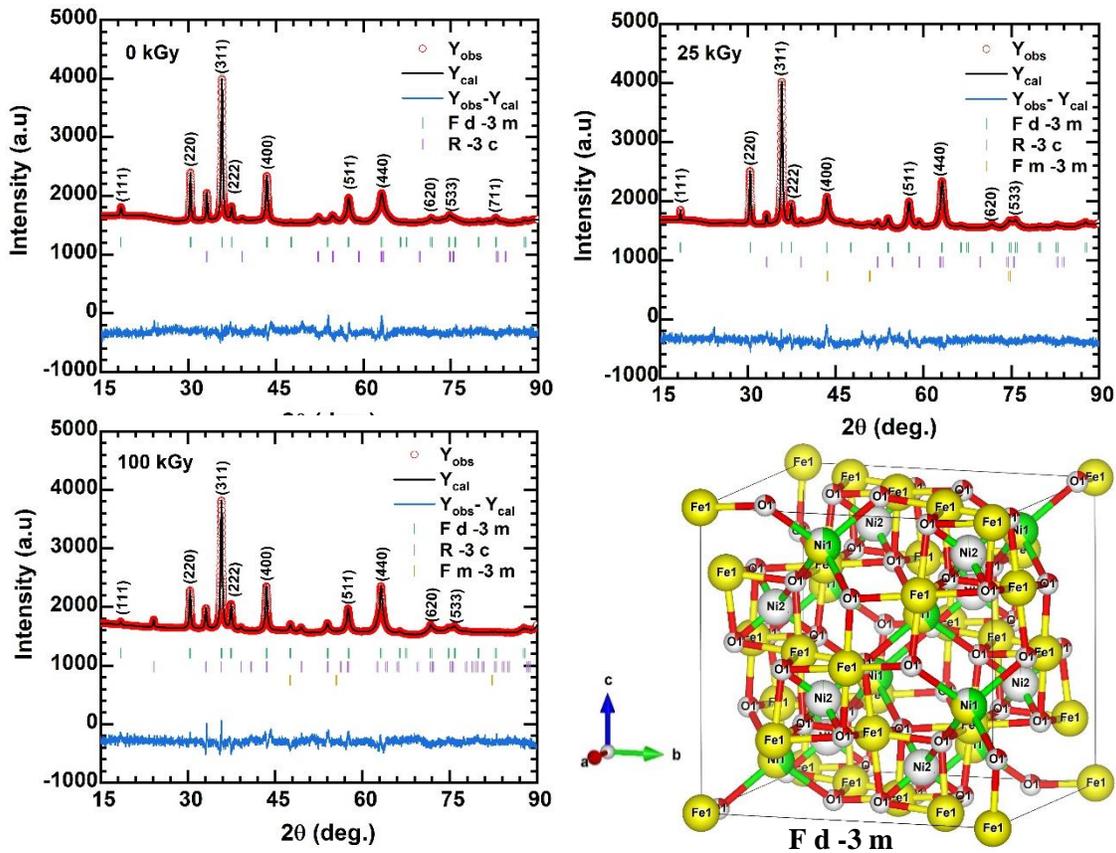

**Fig. 3** The Rietveld refinement of XRD pattern with different γ irradiations and a face centered cubic (fcc) structure of $NiFe_2O_4$

**Table 1** Several structural parameters of pristine and γ-irradiated samples

| Parameters | | | 0 kGy | 25 kGy | 100 kGy |
|---|---|---|---|---|---|
| Lattice constants (Å) | $NiFe_2O_4$ ($a=b=c$) | | 8.3304 | 8.3268 | 8.3189 |
| | $Fe_2O_3$ ($a=b,c$) | | 5.4033, 5.2724 | 5.4022, 5.3045 | 5.0309, 13.7998 |
| | Ni ($a=b=c$) | | --- | 3.5980 | 3.3129 |
| Volume (Å$^3$) | $NiFe_2O_4$ | | 578.1011 | 577.3437 | 575.7062 |
| | $Fe_2O_3$ | | 133.3125 | 134.0625 | 302.4775 |
| | Ni | | --- | 46.5782 | 36.3614 |
| Crystallite size (nm) | $D_{D-S}$ | | 34.04 | 34.71 | 34.12 |
| | $D_{ave.}$ | | 24.55 | 51.16 | 42.61 |
| | $D_{W-H}$ | | 71.23 | 83.23 | 77.78 |
| | $D_{R-R}$ | | 70.12 | 77.78 | 75.01 |
| Rietveld Refinement | $R_{wp}$(%) | | 8.10 | 8.04 | 8.07 |
| | $R_{exp}$ (%) | | 6.95 | 7.04 | 6.98 |
| | G.O.F ($\chi$) | | 1.69 | 1.88 | 1.56 |
| | Oxygen position, (u) | | 0.22920 | 0.25356 | 0.39745 |
| Bond length (Å) | $NiFe_2O_4$ | Fe-O | 1.92552 | 2.11176 | 3.29151 |
| | | Fe-Fe | 2.94526 | 2.94397 | 2.94120 |
| | | $Ni_1$-O | 2.10304 | 1.75146 | --- |
| | | $Ni_2$-O | 1.50415 | 1.85415 | --- |
| | | $Ni_1$-$Ni_2$ | 4.16522 | 4.16340 | 4.15953 |
| | | $Ni_1$-Fe | 3.45362 | 3.45211 | 3.44882 |
| | | $Ni_2$-Fe | 1.80359 | 1.80281 | 1.80110 |
| | $Fe_2O_3$ | Fe-O | 1.4709, 1,5088, 1.5270, 2.2582 | 1.8016, 2.0098 | 1.8154, 2.1871 |
| | | Fe-Fe | 1.5270 | 2.1372 | 3.4068, 2.9507 |
| | $Ni_2$ | Ni-Ni | ---- | 2.5442, 3.5981 | 2.3426, 3.3129 |
| Bond angle (deg.) | $NiFe_2O_4$ | O-Fe-O | 79.28, 100.72, 100.75, 100.73 | 88.40, 91.60, | 9.20, |
| | | O-$Ni_1$-O | 109.45, 109.47, 109. 48 | 109.4712 | 109.4712 |
| | | O-$Ni_2$-O | 109.51, 109.48, 109.45 | 109.4712 | --- |
| | | Fe-O-$Ni_1$ | 117.96 | 126.40 | 116.67 |
| | | Fe-O-$Ni_2$ | 62.03 | 53.60 | O-Fe-$Ni_2$=98.60 |
| | $Fe_2O_3$ | Fe-O-Fe | 61.6417 | 67.9588 | 71.4986 |
| | | O-Fe-O | 85.9783 | 78.7610 | 85.4653 |

The R.R process continues until convergence is reached with values less than 2.0 or close to 1.0, which designates a correlation between the experimental data and adjustment model used. The values of $R_{wp}$, $R_{exp}$, and $\chi^2$ are presented in Table 1. In addition, the estimated Rietveld refined lattice parameter and volume values are also tabulated in Table 1. It can be seen (Table 1) that both the lattice parameter and volume are decreasing simultaneously with increasing the gamma radiations which signifies that atomic position is displaced significantly. The shrinking of lattice parameter and volume due to the lattice vacancies produced after $\gamma$ radiation dose reasons distortion and deviation from the spinel cubic structure. Besides, $\gamma$ radiation typically formed the compressive strain and then generated some disorder into the $NiFe_2O_4$ lattice structure [43].

However, the crystallite size of the samples is calculated from the line width of the (3 1 1) peak using Debye-Scherrer's (D-S) equation [44]: $D_{D-S} = \frac{k\lambda}{\beta cos\theta}$, where $'D_{D-S}'$ is the crystallite size, $'\lambda'$ is the wavelength of CuKα radiation (1.5404Å), 'β' is the full width half maxima (FWHM), and 'θ' is the diffraction angle of the strongest characteristic peak. After that, we have also estimated the average crystallite size ($D_{ave.}$) using the Lorentz function by Origin pro-2018. From Table 1, it is clearly evident that in both cases the $D_{D-S}$ and $D_{ave.}$ increases drastically at 0 and 25 kGy samples, and decreases at the 100 kGy sample. To get the more accurate crystallographic parameters such as crystallite size, strain, atomic structure, bond length and bond angle, the Williamson–Hall (W-H) method and Rietveld refinement (R.R) is performed precisely. Initially, the W-H method gives the values of crystallite size and strain concurrently.

The equation of W-H method is [45]: $\beta cos\theta = 4\varepsilon_{WH} sin\theta + \frac{K\lambda}{D_{WH}}$, where is an intercept in W-H plot that corresponds to the crystallite size (DW−H) and slope $\varepsilon_{W-H}$ corresponds to the strain as shown in Fig. 4. It can be seen that $D_{W-H}$ lies between 71.23 to 83.23 nm which emulates the $D_{D-S}$ and $D_{ave}$ trend. Additionally, the Rietveld refined (R-R) crystallite size is found from 70.12 to 77.78 nm. It is observed that the crystallite size increases after 25 kGy γ radiation doses, which is ascribed to the following: γ radiation interacts with the particles or grains and ionizes $Fe^{3+}$ ions to $Fe^{2+}$ ions [46] as follows:

$$Fe^{3+} + \gamma \rightarrow Fe^{2+} + e^- \quad (1)$$

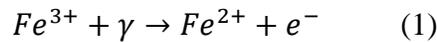

But the crystallite size decreases at 100 kGy γ radiation dose due to the modification of the ratio of $Fe^{2+}/Fe^{3+}$ ions in octahedral-B site that can be expressed by the following equation [47]:

$$Fe^{2+} + \gamma \rightarrow Fe^{3+} + e^- \quad (2)$$

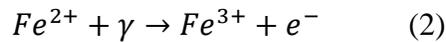

Through the R-R analysis, we have determined the bond length and bond angle to know the gamma radiation effect into the $NiFe_2O_4$ structure. Firstly, in the case of $NiFe_2O_4$, the Fe-O and $Ni_2$-O bond length is grown up abruptly while Fe-Fe, $Ni_1$-O, $Ni_1$- $Ni_2$, $Ni_1$-Fe and $Ni_2$-Fe bond distance declining after using gamma radiations as shown in Table 1 and Fig. 5.

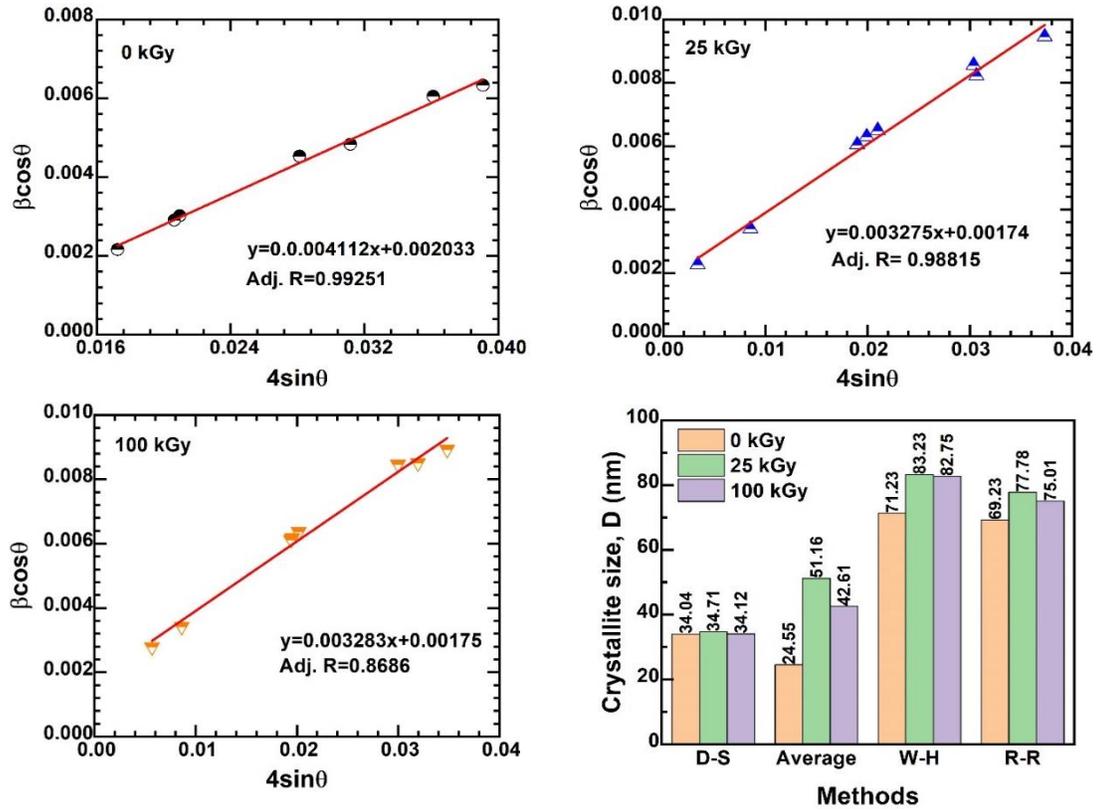

**Fig. 4** Williamson–Hall plot for NiFe$_2$O$_4$ with different γ irradiations. A comparison crystallite size studies among various methods.

Interestingly, there is no bond length appeared between Ni$_1$-O and Ni$_2$-O at 100 kGy sample that may be attributed to the development of Fe-O bond distance. Further, multiple bond distance of Fe-O, Fe-Fe and Ni-Ni is detected for Fe$_2$O$_3$ and Ni structures as shown Table 1. Likewise, in the event of a bond angle, we have found multiple bond angles in the NiFe$_2$O$_4$ and Fe$_2$O$_4$ structures. The O-Ni$_1$-O and O-Ni$_2$-O bond angle almost analogous, on the contrary, Fe-O-Ni$_1$ and Fe-O-Ni$_2$ bond angle completely different in the NiFe$_2$O$_4$ structure. What's more, an interest point has build that Ni$_2$ atom does not make bond length and bond angle as well at 100 kGy sample. It can be speculated that owing to the higher gamma radiation Ni$_2$-O bond split up and then make a new bond only with the Fe atom as shown in Fig. 5.

However, the Hopping length (distance between magnetic ions in tetrahedral site-$L_A$ and octahedral sites-$L_B$) is another important structural parameter which estimated by these equations [48]: $L_A = \frac{1}{4}a\sqrt{3}$ and $L_B = \frac{1}{4}a\sqrt{2}$, where $'a'$ is the lattice parameter of NiFe$_2$O$_4$ structure. The calculated values of the $L_A$ and $L_B$ are presented in the Table 2. It is seen (Table 2) that the hopping lengths in $L_A$ and $L_B$ sites decrease concurrently with changing gamma radiations. But other groups showed that in the case of ZnFe$_2$O$_4$ and CoFe$_2$O$_4$, hopping lengths increased after using gamma radiations as they have found greater values of lattice parameters [19][29].

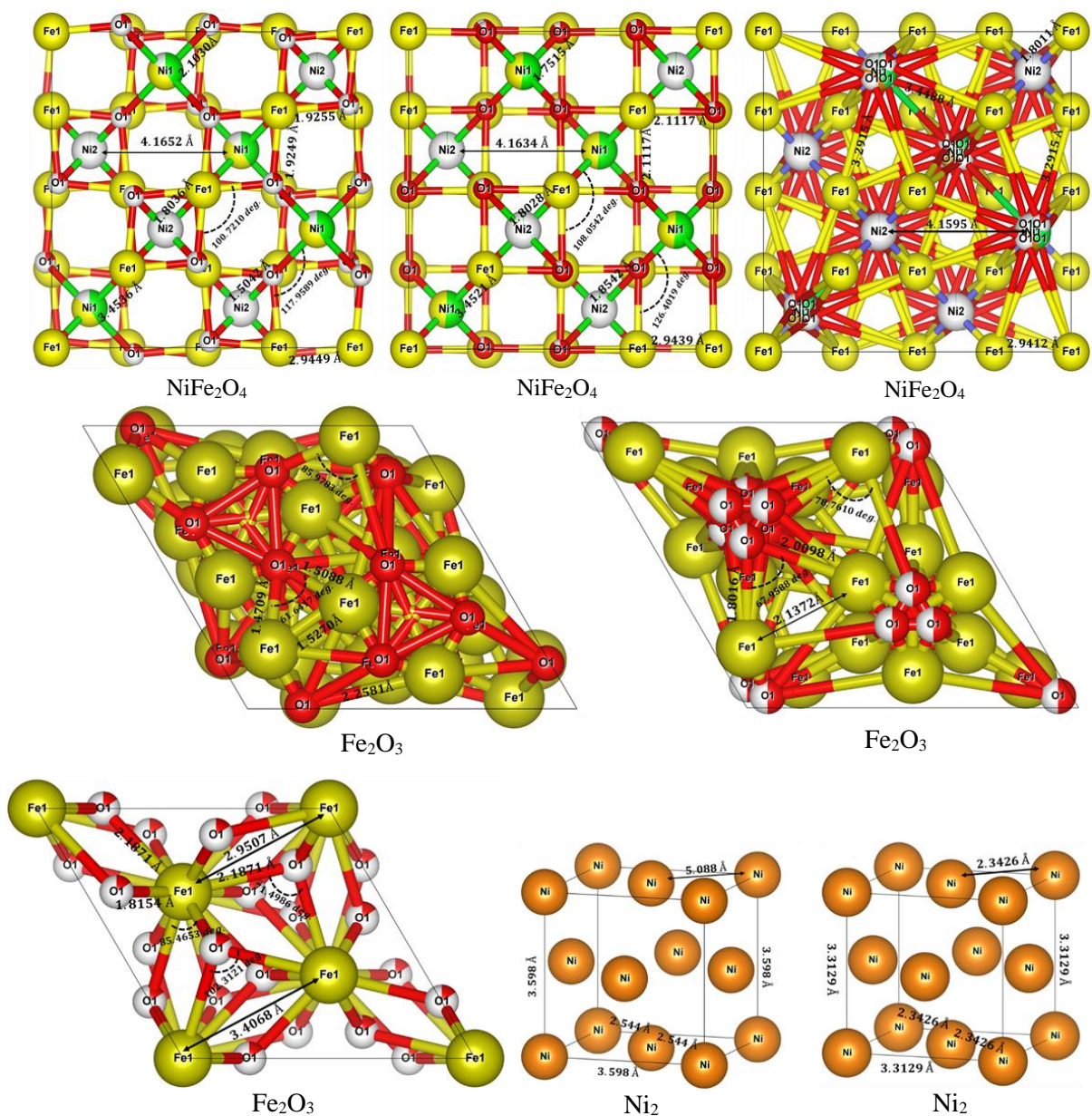

**Fig. 5** Several Rietveld refined structures with denoting bond lengths and bond angles.

In the present study, the lattice parameter, volume, and bond length are shrunk with altering gamma radiations which are firmly confirmed that these structural parameters control the Hopping lengths. Further, the inter-ionic distances (average bond lengths) at tetrahedral-$R_A$ and octahedral sites-$R_B$ is calculated using the following relations [49]:

$$R_A = a\sqrt{3}\left(u - \frac{1}{4}\right) \quad (3)$$

$$R_B = a\sqrt{3u^2 - 2u + \frac{1}{16}} \quad (4)$$

Where 'u' represents the Rietveld refined oxygen positional parameter which values are 0.22920, 0.25356 and 0.39745 for 0, 25 and 100 kGy gamma radiation samples, respectively. Our refined 'u' values are analogous with previous reported data [50]. But other groups have been taken theoretical value of oxygen position parameter (u=0.381Å). Patange et al. has chosen 'u' value for $Al^{3+}$ substituted $NiFe_2O_4$ nanoparticles despite the Rietveld refinement analysis [51]. It is observed that 'u' is increased with rising gamma radiations, and $R_A$ is increased while declining $R_B$ as shown in Table 2.

**Table 2** The estimated Hopping lengths and average bond length of $NiFe_2O_4$ for various γ irradiations

| Parameters | | 0 kGy | 25 kGy | 100 kGy |
|---|---|---|---|---|
| Hopping length (Å) | $L_a$ | 3.6072 | 3.6056 | 3.6022 |
|  | $L_b$ | 2.9445 | 2.9440 | 2.9412 |
| Average bond length (Å) | Tetrahedral site ($R_A$) | 1.5035 | 1.8542 | 3.9257 |
|  | Octahedral sites ($R_B$) | 2.2691 | 2.0525 | 1.9331 |

The change of $R_A$ and $R_B$ can be explained on the basis of the drastic movement of oxygen position parameters. Since oxygen position is developed considerably with rising gamma radiations, so the $NiFe_2O_4$ structure becomes slowly close to the fcc structure and oxygen ions are moving towards octahedral coordinated at the end. However, several studies are investigated on spinel ferrite by gamma doses. Here, we have reported on the effect of gamma dose over several spinel ferrites in Table 3.

**Table 3** A comparison study for various spinel ferrites under γ irradiations

| Sample | Rietveld refinement | No. of phases | Lattice constant | Determined bond length and bond angle | Determined $R_A$ and $R_B$ | Hopping length | Ref |
|---|---|---|---|---|---|---|---|
| $ZnFe_2O_4$ | No | Single | Increase | No | No | Increase | [19] |
| $CoFe_2O_4$ | No | Single | Increase | No | No | Increase | [29] |
| $NiFe_2O_4$ | Yes | Three | Decrease | Yes | Yes | Decrease | present |

### 3.3 Optical properties

The optical band gap ($E_g$) of the γ pristine and γ-irradiated $NiFe_2O_4$ samples has been calculated from the Kubelka–Munk (K–M) function through the following equation [52]:

$$F(R) \propto \alpha = \frac{(h\vartheta - E_g)^2}{h\vartheta} \quad (5)$$

Where, 'R' is the reflectance, $'\alpha'$ is the absorption coefficient, $'h\vartheta'$ is the incident light energy and $'E_g'$ is the optical band gap. Fig. 6 shows the $E_g$ plots of the $NiFe_2O_4$ treated with different γ irradiation doses. The $E_g$ is calculated by plotting a graph between $(F(R) \times h\vartheta)^2$ versus $h\vartheta$. In the pristine $NiFe_2O_4$, the $E_g$ is 1.85 eV. Subsequently, for 25 and 100 kGy irradiation doses, the $E_g$ is found to be 1.80 and 1.89 eV, respectively. Thus, we observe that the $E_g$ shrinkages when low doses of γ irradiation are used, while with high dose γ irradiation the $E_g$ is increased. The change in $E_g$ can be expounded by the following ways: (i) quantum-size effect may be accountable for altering the $E_g$. Because it is resulting from the strong interaction between the surface of $NiFe_2O_4$ and γ photons [53].

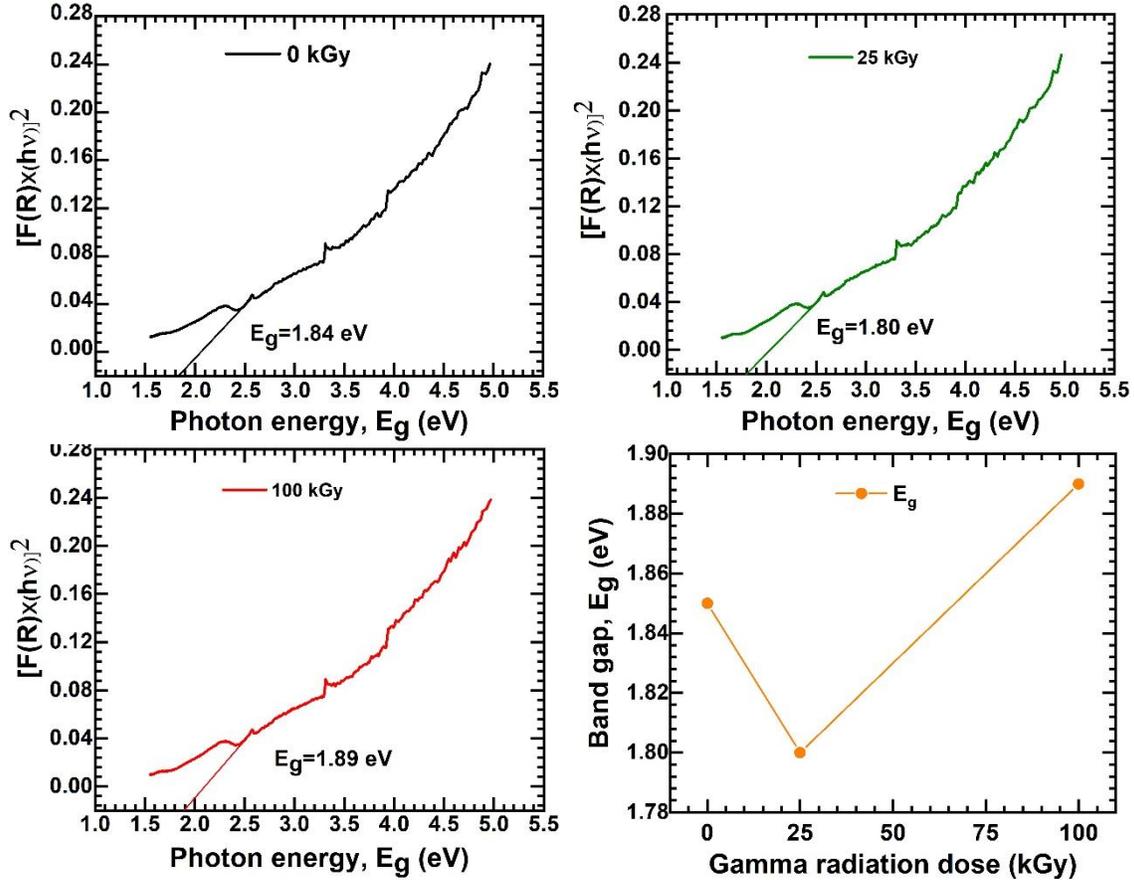

**Fig. 6** The $E_g$ of $NiFe_2O_4$ with various γ irradiation doses and with that a comparison of $E_g$ study

(ii) decrease of $E_g$ may be ascribed to the creation of localized states into the $NiFe_2O_4$ structure due to structural defects [54]. (iii) enhancement of $E_g$ due to the optical scattering at the particle or grain boundaries and intrinsic absorption [21]. Besides, it is anticipated that the $\ddot{V}_O$ is created within $NiFe_2O_4$ structure during higher dose of γ irradiation that will play a significant role to increase the $E_g$ [55]. However, it is manifested that the γ irradiation significantly affects the crystallite size and hence alter the optical band gap of $NiFe_2O_4$.

### 3.4 FTIR analysis

Fig. 7 displays the FTIR diagram of pristine $NiFe_2O_4$ and $NiFe_2O_4$ irradiated with γ-doses for analysis of chemical bonds within the resulted products. The prominent bands with sharp peaks have been observed at 365 cm$^{-1}$ and 547 cm$^{-1}$ which are assigned to stretching vibrations of Ni-O bond in octahedral complexes and tetrahedral Fe-O vibrations, respectively [56][57]. This analysis of FTIR presents the consistency with published results in numerous bodies of literature

[58][59][60] and ensures the construction of the spinel $NiFe_2O_4$ nanoparticles identifying characteristic peaks in FTIR diagram. The FTIR studies clearly support the formation of Ni ferrite as observed in XRD analysis. The presence of band positions in the ranges of 1000–1300 cm$^{-1}$ and 2000–3000 cm$^{-1}$ demonstrates the survival of O–H, C–O, and C=H stretching mode of organic compounds [61]. Two peaks appear at 1381 cm$^{-1}$ and 3754 cm$^{-1}$ in case of the pristine and γ-irradiated products, which belongs to stretching vibration of C–H and the contribution of N–H bonds, respectively [61].

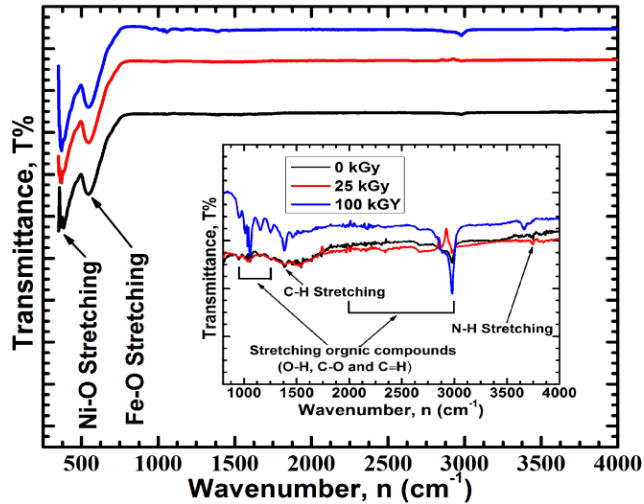

**Fig. 7** FTIR spectra of pristine and γ-irradiated $NiFe_2O_4$

### 3.5 Magnetic properties

Fig. 8 shows the room temperature magnetization (M) for the applied field (H) of as-prepared $NiFe_2O_4$ nanoparticles before (0 kGy) and after irradiation (25 kGy and 100 kGy). The applied field was ranged from -20 kOe to +20 kOe, and it is apparent that the magnetization is not saturated at 20kOe. Low coercivity and low hysteresis indicates the ferrimagnetic behavior of the samples. The magnetic parameters such as saturation magnetization ($M_s$), Remanence ($M_r$) and coercivity ($H_c$) are summarized in Table 4. The saturation magnetization ($M_s$) was found for the pristine sample was ~ 28 emu/g, which is less than the bulk $NiFe_2O_4$. It can be attributed to the fact that $NiFe_2O_4$ becomes mixed spinel from the inverse spinel structure at the nanoscale [62].

The tetrahedral A-site consists of ferromagnetically ordered $Fe^{3+}$ ions, and the octahedral B-site consists of $Fe^{3+}$ and $Ni^{2+}$ ions. The magnetic contribution in the inverse spinel arises from the $Ni^{2+}$ in the octahedral B-site. Due to antiferromagnetic ordering, the $Fe^{3+}$ moments from both the A-site and B-site cancel each other. Considering the magnetic moments of $Fe^{3+}$ and $Ni^{2+}$ ions are 5 $\mu_B$ and 2 $\mu_B$, the net magnetic moment is 2 $\mu_B$ by Neel's two sublattice model [63]. Therefore, the proposed change in the cation distribution at the nanoscale can be interpreted as $(Ni_xFe_{1-x})_A[Ni_{1-x}Fe_{1+x}]_B$. However, the presence of $Ni^{2+}$ might not be that dominant in the tetrahedral A-site for the pristine samples. Also, the synthesis methods significantly impact the magnetic properties on spinel ferrites [64].

There is an increment in the $M_s$ from 28 emu/g to 41 emu/g after the $\gamma-$irradiation of 25 kGy. After the $\gamma-$irradiation, the degree of mixed-phase may have increased, and more $Ni^{2+}$ has transferred to the A-site. Also, the irradiation releases the pinned domains on the surface, and the significant increment of magnetization was observed [65][66]. Furthermore, the onset of crystallite size growth was observed after the irradiation. Crystallite size growth may accompany the migration of $Fe^{3+}$ ions to the B-site [2][67]. The decreasing trend in the lattice parameter also established the emergence of strong ionic interactions among the lattice sites. Therefore, the enhanced saturation magnetization was observed.

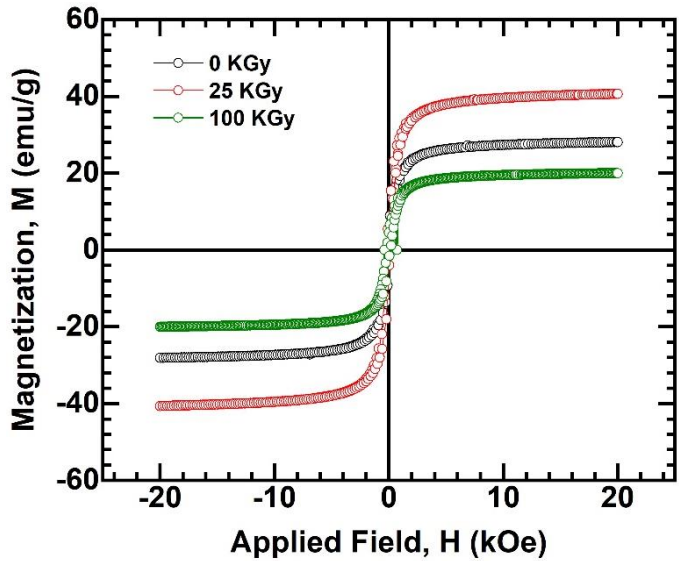

**Fig. 8** M-H curves of the pristine and $\gamma-$irradiated samples of $NiFe_2O_4$.

Consequently, it will be interesting to investigate the magnetic properties on exposing the samples at higher $\gamma-$radiation. However, a strong $\gamma-$irradiation of 100 kGy decreased the $M_s$ drastically to 20 emu/g. The strong $\gamma-$irradiation causes the adverse effect on magnetic ordering by the ion-induced disorder. Also, high energy ions may penetrate the sample. The host atoms and molecules interact with the irradiated gamma photons via inelastic collision.

**Table 4** Magnetic parameters of pristine and γ-irradiated samples of $NiFe_2O_4$

| Radiation (kGy) | $M_s$(Ex) (emu/g) | Remanence, $M_r$(emu/g) | Coercivity, $H_c$ (Oe) |
|---|---|---|---|
| 0 | 28.08 | 4.87 | 0.0022 |
| 25 | 40.63 | 8.44 | 0.0303 |
| 100 | 20.00 | 4.87 | 0.0216 |

The interaction causes energy loss and introduces defects or partial amorphization depending on the amount of energy lost [68]. Hence, a decrease in crystallite size was observed. The reduction of crystallite size introduces spin canting, magnetic dead layers and weakening of super-exchange interaction. Consequently, a substantial reduction in saturation magnetization has emerged.

## 4 Conclusions

Nanoparticles $NiFe_2O_4$ ferrites in spinel cubic structure have been synthesized via sol-gel auto-combustion route. We have investigated the structural, morphological, magnetic and optical properties of the resultant ferrite nanoparticles with the variation of γ-doses. FESEM micrographs clearly indicate the variation in morphology and aggregation nature within the bare and γ-irradiated Ni ferrite products. According to the XRD investigation of the conducted research, it is concluded that an irregular variation in crystallite size leads from the results of Scherrer method. In addition, analysis of the W-H plot and R-R exhibit the identical trend of variation upon incorporation of γ-irradiation into the $NiFe_2O_4$ crystal network. The crystallite size of pristine Ni ferrite sample is observed as 24.55 nm-71.23 nm. The result after γ-irradiation with 25 kGy shows the increment nature of crystallite size and then decrease after exposure to high γ-dose (100 kGy). From the Rietveld analysis numerous structural parameters such as bond length, bond angle, hopping length etc. are estimated in this current study. The optical bandgap energy is estimated at 1.85 eV in the pristine product and consequently, is the case of low (25kGy) and high (100 kGy) γ-doses it corresponds to be 1.80 and 1.89 eV, respectively. Two absorption bands located at 365 $cm^{-1}$ and 547 $cm^{-1}$ have been detected from the FTIR measurement which indicates the stretching vibrations at the octahedral site of Ni-O and tetrahedral Fe-O vibrations, respectively. This obtained observation confirms the formation of spinel cubic phase in both pristine and γ-irradiated Ni ferrites nanoparticles supported by XRD results. PPMS measurement reveals that the $M_S$ value for pristine product is noticed as 28.08 emu/g. A considerable increase in saturation magnetization from 28.08 emu/g to 40.63 emu/g in low γ-irradiated compound (25 kGy) which may be due to the surface canting effect affected by the agglomeration nature of ferrite particle and then decreasing nature after exposure to high γ-dose.